\documentclass[twocolumn,showpacs,preprintnumbers,amsmath,amssymb,floatfix]{revtex4}
\usepackage{graphicx}

\begin{document}

\title{\center{Fabrication of metal nanowires}}
\author{D. Natelson\\
Dept. of Physics \& Astronomy and Electrical \& Computer Engineering\\
Rice University\\
Houston, TX 77005}



\begin{abstract}

  This review article discusses and compares various techniques for
fabricating metal nanowires.  We begin by defining what we mean by a
nanowire, and why such nanostructures are of scientific and
technological interest.  We then present different fabrication
methodologies, describing in some detail the advantages of each.
``Top-down" techniques discussed include: electron beam lithography;
scanned probe lithography; step-edge and molecular beam
epitaxy templating; and nanotube templating.  ``Bottom-up"
methodologies covered include: electrodeposition into etched porous
media; direct chemical synthesis; step-edge decoration; and
strain-mediated self-assembly.  We conclude by summarizing our
observations and briefly discuss the future of metal nanowire
fabrication.

\end{abstract}

\maketitle

\section{\small{INTRODUCTION}}
\label{sec:intro}

  Colloquially, a wire is a metallic structure extended in one
(longitudinal) direction and strongly confined in the other two
(transverse) directions.  A fundamental attribute of such a
two-terminal device is electrical continuity, {\it i.e.} the ability
to transport charge along its length under a longitudinal potential
bias.  There is much scientific and technological interest in
fabricating and understanding metal wires with transverse dimensions
approaching the nanometer scale.  We call such structures {\it
nanowires}, and this article reviews a number of techniques for the
fabrication of these objects.

For our purposes a metal nanowire has transverse dimensions ($w,t$)
substantially below 100~nm, an aspect ratio significantly greater than
one ($L >> w,t$), and is composed of a material that is metallic in
the bulk ({\it e.g.} gold, copper, nickel).  This definition is
deliberately constructed to exclude from discussion some other
structures known colloquially as nanowires: {\it e.g.}, nanoscale metal
point contacts\cite{Ruitenbeek99Springer,YansonetAl98Nature}; extended
semiconductor crystals\cite{CuietAl01APL} and heterostructures
(so-called ``quantum wires")\cite{PfeifferetAl93JCG,LiangetAl99APL};
and molecules used as two-terminal conductors
\cite{TansetAl97Nature,Tour00ACR}.

  It is useful to compare the sizes of nanowires
with some of the physically significant length scales in
metals.  These comparisons emphasize the difference between 
nanowires and the other elongated structures mentioned above, and also
show why the sub-100~nm transverse scale is interesting.

  The Fermi wavelength, $\lambda_{\rm F}$, is one relevant length
scale.  In typical bulk metals the Fermi wavelength associated with
the conduction electron Bloch waves is on the order of 0.1~nm.  This
short wavelength is due to the high spatial density of electrons.
Since $\lambda_{\rm F} << w,t$ except for wires of atomic
cross-section, most nanowires under discussion have {\it many}
conducting channels, and are well-described by three dimensional Fermi
Liquid Theory.  More exotic physics ({\it e.g.} Luttinger liquid
properties) only becomes relevant in extremely narrow structures, when
$w,t \rightarrow \lambda_{\rm F}$.  Note that in contrast,
$\lambda_{\rm F}$ in doped semiconductor nanowires may be on the order
of tens of nanometers.

  Another closely related quantity is the screening length, $L_{\rm
sc}$, the scale over which electrostatic impurities are screened by
the conduction electrons.  This is on the same order as $\lambda_{\rm
F}$.  As transverse wire dimensions are reduced, an increasing
fraction of the metal atoms are within $L_{\rm sc}$ of the wire
surface.  

  Other relevant sizes are $\ell$, the electronic mean free path for
elastic scattering, and $\ell_{\rm tot}$, the mean free path for
either inelastic or elastic scattering.  A ``good'' metal has
$\ell/\lambda_{\rm F} >> 1$; that is, electronic partial waves
propagate many wavelengths between scattering events.  Since metal
nanowires are often nanocrystalline, elastic scattering in such
structures is strongly influenced by grain boundary scattering.
Similarly, if the transverse dimensions of the wire are sufficiently
small ($\ell \sim w,t$) then boundary scattering can also strongly
influence nanowire conduction properties.  This limit is certainly
achievable at low temperatures for ``clean'' metals.

  Related to the inelastic mean free path is the coherence length,
$L_{\phi}$.  Consider introducing an electron in a single-particle
eigenstate of the wire, and then allowing the electron to propagate
while undergoing interactions with other dynamic degrees of freedom
({\it e.g.} phonons, photons, magnetic impurities, other electrons).
$L_{\phi}$ is the distance traversed before the phase of the electron
partial wave becomes poorly defined through interactions with the
environment.  Quantum interference corrections to electronic
conduction (often referred to as ``mesoscopic''
phenomena\cite{Datta95,Imry97}) are typically relevant on this length
scale.  Because of the temperature dependence of processes involving
those dynamic degrees of freedom, $L_{\phi}$ often varies strongly with
$T$, tending toward larger values at lower temperatures.  In a metal
like silver at 1~K, $L_{\phi} \sim 1~\mu$m.

Once $L_{\phi}>w,t$, the wire is one-dimensional with respect to
quantum phase coherence.  One important property of such
one-dimensional structures is that the motion of a single impurity or
defect will then affect the conducting properties of the entire wire.
Every trajectory that brings an electron from one end of the wire to
the other will have to pass through the same coherent volume as the
defect in question.  Similarly, suppose some charged defect adjacent
to the nanowire changes its state.  Classically this should only
affect electron trajectories passing within $L_{\rm sc}$ of the
nanowire surface; however, because of quantum coherence and mixing of
transverse channels, all trajectories passing within $L_{\phi}$ must
now be considered.

As wire transverse dimensions are reduced, quantum corrections to 
conduction become relevant at higher temperatures.  In the limit of
atomic-scale metal contacts, quantum effects dominate electronic
transport even at room temperature\cite{Ruitenbeek99Springer}.  In
some of the smallest nanowires considered in this article, conductance
fluctuation noise from quantum interference and scatterer
motion\cite{Feng91,Giordano91} is clearly detectable at temperatures
above that of liquid nitrogen\cite{NatelsonetAl00SSC}.

Finally, one more useful quantity is the thermal length $L_{T} =
\sqrt{ \hbar D/k_{\rm B}T }$ in diffusive wires and $\hbar v_{\rm
F}/k_{\rm B}T$ in ballistic wires, where $v_{\rm F}$ is the Fermi
velocity.  Consider starting out two electrons near the Fermi surface
with energies differing by $k_{\rm B}T$, but beginning on the same
trajectory.  The length $L_{T}$ is the distance that the electrons
would typically travel before their partial waves are out of step 
due to their slightly different energies.  This scale also determines
the thermal smearing of the conductance fluctuation noise described
above. 

Metal nanowires with transverse dimensions well below 100~nm are of
scientific interest because this size range spans the lengths
mentioned above.  Surface scattering effects, changes in screening,
and the increased importance of quantum corrections to the conductance
are all topics of research that may be addressed through novel
nanowire samples.  Moreover, metals with substantial electronic
correlation effects such as ferromagnetism and superconductivity
possess additional length scales ({\it e.g.} domain wall thickness,
superconducting coherence length) that further complicate wire
properties.  Nanowires are ideal tools for the study of such
physics.

With the continued trend toward further technological miniaturization,
nanowires are likely to be industrially relevant as well.  Already
the physical gate length for transistors is predicted to be
as small as 25~nm by 2007\cite{ITRS01}.  With the possibility
of electronic devices based on individual molecules being 
seriously explored, fabrication of metal nanostructures on 
comparable scales is a topic of much interest.

  As has been pointed out by other authors\cite{XiaetAl99CR},
fabrication processes at these length scales fall into three broad
categories: {\it lithographic} processes (Sec.~\ref{sec:litho})
involve pattern definition through a drawing step; while {\it
templating} approaches (Sec.~\ref{sec:template}) use alternative means
to produce a nanodetailed template and replicate some feature of that
template in metal; and finally {\it chemical synthesis and
self-assembly} may be employed without any patterning step at all.  We
begin by considering lithographic techniques.

\section{\small{LITHOGRAPHIC TECHNIQUES}}
\label{sec:litho}

  Lithographic techniques for nanowire fabrication may be divided into
two classes: additive and subtractive methods.  In many additive
techniques some form of {\it resist} is applied to the entire
substrate; the resist is then modified in a {\it pattern definition}
step, turning the resist layer into a stencil, with resist material
removed to expose the underlying substrate.  Metallization then
occurs, in which the entire resist layer is coated with a metal film
of thickness $t$, usually by evaporation or sputtering.  The final
step is {\it liftoff}, in which the remaining resist is removed by
chemical means, leaving behind metal on the substrate only where the
resist had been patterned.  Alternately, resist-free techniques may be
employed, in which the metal is ``drawn" directly onto the substrate.

  Subtractive techniques begin by metallizing the entire substrate
with the desired wire material.  Some form of resist may then be
applied and patterned, with resist remaining where the wire is desired
to be formed.  Unwanted metal is then removed by physical etching or chemical
means.  Any remaining resist is then lifted off
chemically.  Again, there are also resist-free approaches that locally
remove the excess metal to form the nanowires.  For nanowire
fabrication, subtractive lithographic approaches are less common than
additive ones.

\subsection{Photolithography}

  We only briefly consider photolithography because of its size
limitations.  For more complete discussions of this extremely
wide-spread technique, see Refs. \cite{Moreau88,Wong01}.  In this
technique chemical changes in a resist material are photoactivated,
with some sort of optical mask used to spatially define the regions of
photochemical activity.  Optics between the mask and the substrate are
used to further reduce mask feature sizes when exposing the resist.

An intrinsic limitation of this technique is the diffraction limit of
the wavelength of light involved.  Current industrial practice typically uses
248~nm light, and through clever mask designs and specially tailored
polymeric resists\cite{ReichmanisetAl01MS}, feature sizes below 100~nm
are possible.  Further reductions are achievable, but there are
enormous engineering challenges involved, since wavelengths in the
deep ultraviolet (DUV) and x-ray require significant changes in the
optical elements used in patterning.  Extensions of projection
photolithography to the 10~nm transverse size scale seem unlikely to
occur, except perhaps through scanning near-field
methods\cite{BetzigetAl92Science,KuwaharaetAl00ME}.  We do not
consider nanowire production by photolithography further.

\subsection{Electron-beam lithography}
\label{sec:EBL}

  One of the most flexible and widely used techniques for producing
narrow metal nanowires is electron beam lithography (EBL).  We
consider this method in some detail, and certain steps ({\it e.g.}
metallization) are of general concern to the other techniques that we
discuss.  

\subsubsection{Patterning}
In EBL, a resist of some type (often polymeric) is applied to the
substrate, and a focused beam of high energy (typically $>$ 30~kV)
electrons is scanned over the surface, tracing out the desired
pattern.  The beam hits the resist and locally ionizes the constituent
atoms, breaking chemical bonds and producing a cloud of secondary
electrons, many of which are also sufficiently energetic to do further
damage to the resist.

 In ``positive" polymer resists, exposed polymers are broken down into
smaller units, typically with enhanced solubility in some developer
solution when compared to the original material\cite{Moreau88,Wong01}.
Development removes the exposed resist, with the remaining material
comprising a stencil for further processing.  ``Negative" resists are
less common; in these materials the electron beam locally crosslinks
the polymer, so that development removes the {\it un}exposed resist.
Positive resists nearly always have superior resolution for isolated
features such as single nanowires, in part because of some techniques
described below.  Thus we confine the remainder of our EBL discussion
to positive resists.

Polymethylmethacrylate (PMMA) is the most commonly employed positive
e-beam resist; different molecular weights of polymer (typically
495~kDaltons and 950 kDaltons) and different concentrations of polymer
in carrier solvent (typically anisol or chlorobenzene) are used to
tailor the resist's exposure and viscosity properties.  The resist is
commonly spun onto a planar substrate which is then baked for some
time.  The bake is at a temperature high enough to drive off the
carrier solvent yet low enough to avoid crosslinking or decomposing
the polymer (say 160$^{\circ}$C for 2 hours).

Nanowire production with a positive resist requires pattern definition
in the resist, development that preserves that pattern, and a
successful metallization and liftoff step.  Each of these steps
contains subtleties that determine the size limits of EBL-produced
nanowires.

\begin{figure}[h!]
\includegraphics[width=8.5cm, clip]{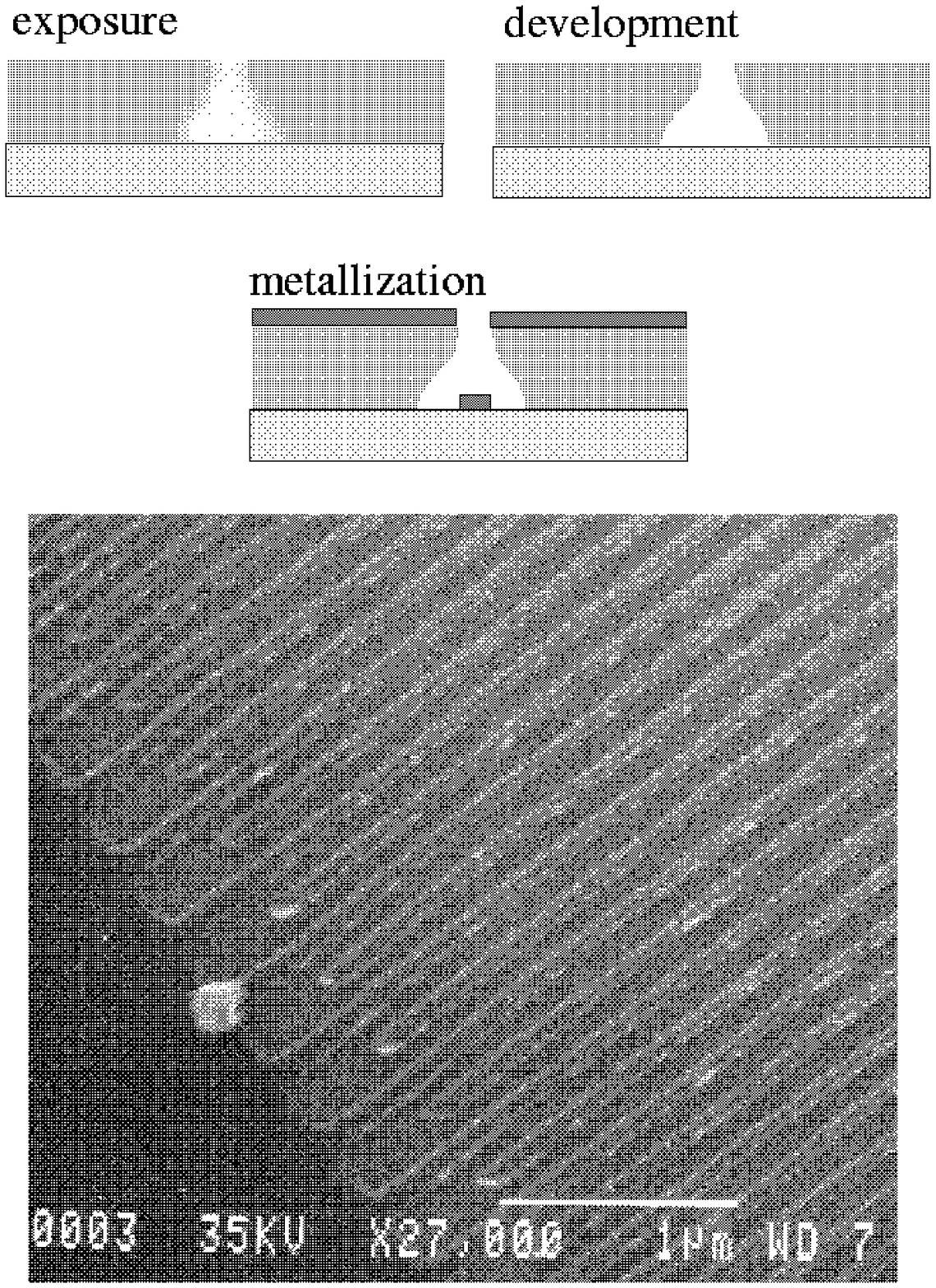}
\caption{(a) Cross-section of ideal resist exposure, showing resultant
development and metallization.  The larger exposure near the substrate
in the top image is due to secondary electrons produced in the
substrate.  (b) An example of an array of interdigitated Al nanowires
made by e-beam lithography and liftoff processing.  Image courtesy
R.L. Willett, Bell Labs.}
\label{fig:liftoff}
\end{figure}

Electron wavelength is usually not an issue, since sub-{\AA}ngstrom
wavelengths are achievable with beam voltages in excess of a
few kV.  Similarly, electron beam spot size is generally not a
limiting factor; electron optics can produce electron beams with
atomic resolution\cite{KrivaneketAl99U}.  The exposed area of resist
is primarily limited by the size of the secondary electron cloud,
which is usually much larger than the impinging beam size.  All other
things being equal, higher column voltages tend to produce narrower
exposed patterns for two reasons: lower beam transit time to the
sample minimizes relative jitter between the gun and the sample; and
higher incident electron momentum produces a secondary cloud farther
into the substrate, with fewer secondary exposures within the resist.
A similar secondary distribution in the resist may be achieved by
using extremely thin samples to reduce the total secondary
yield\cite{SmithetAl90PT}.

Figure \ref{fig:liftoff}a illustrates an ideal resist exposure profile
for liftoff metallization.  When developed, this exposure profile
should leave a clean resist overhang for liftoff, as shown.  Other
approaches for obtaining such a profile include double-layer resist
schemes, with the lower layer possessing a higher
sensitivity\cite{GrobmanetAl79IEEE,RooksetAl87JVSTB}, and using an
extremely thin metal film as a top layer to produce a large downward
secondary yield\cite{Sharifi96}.  More exotic schemes include
inorganic resists such as CaF$_{2}$\cite{HarrisonetAl82APL} and
self-assembled monolayer (SAM) resists\cite{TiberioetAl93APL}.

It is challenging to convert even an ideal resist exposure profile
into a completed nanowire at deep-sub-100~nm scales.  
Beam jitter must be extremely well-controlled
to avoid edge roughness.  The
development process also plays a crucial role.  Significant
complications arise in the development process, including: the 
transport of developer into and out of the exposed resist; 
the strong intermolecular forces between development products and
unexposed resist at sub-10~nm scales; and resist
swelling\cite{ChenetAl93APL,YasinetAl01APL}.  These difficulties 
are enhanced as the aspect ratio of the resist exposure profile is 
increased, so that thinner resist layers are generally used for 
higher resolution lithography. 

An alternative approach is to make do with larger feature sizes
in the original resist pattern, and use subsequent steps to 
reduce the linewidth.  For example, by applying SAM coatings to
parallel metal leads, the gap between the leads can be reduced
by precise amounts; subsequently evaporated metal
can then form nanowires with widths $\sim$12~nm, significantly
narrower than the original pattern linewidth\cite{HatzoretAl01Science}.

\subsubsection{Metallization}

Finally, as we shall see again in other fabrication
techniques, the detailed choice of metallization material and method
can strongly affect final nanowire morphology.  Evaporation (thermal
or electron beam) and sputtering are common metal deposition methods.
The sticking and diffusion properties of the metal on the substrate
are crucial to forming uniform and continuous metal films, let alone
nanowires.  For example, on clean, room-temperature GaAs, pure Au is
highly mobile, adhering better to itself than to the substrate.
As a result, a Au film nominally a few nm thick tends to break up into
discrete grains rather than to form a continuous layer of uniform
thickness.

One metallization approach that has been successful in a number of
nanowire fabrication methods is to minimize the grain size of the
metal.  For example, alloys of Au and Pd are known to have reasonably
good adhesion qualities and grain sizes on the nm
scale\cite{ProberetAl80APL,Giordano80PRB}.  The ultimate limit of
small grains is an amorphous metal.  Quench condensation by
evaporation onto extremely cold substrates can produce continuous and
films of a number of metals at few-monolayer coverages
\cite{DynesetAl78PRL,HerzogetAl96PRL,ButkoetAl00PRL}.  Another
approach is to use ``naturally'' amorphous metals, such as MoGe
alloys\cite{GraybealetAl84PRB}.  Even with atomic lateral definition
of a pattern, the narrowest nanowires possible are determined by the
surface physics of the metal and any metal-substrate interface.

To summarize: EBL is a very flexible nanofabrication technique
capable, under the best circumstances, of sub-10~nm linewidths on a
variety of substrates.  With more typical conditions, nanowires with
widths on the order of 20-40~nm may be produced fairly routinely.
Edge roughness is a significant issue, and even with ideal liftoff
lateral definition is influenced by the patterning beam itself, the
development process, and the morphology of the metal.  A major
advantage of the technique is that nanowires may be 
positioned in a variety of configurations and directions on
the substrate.  Furthermore, EBL requires no exotic contraints on
sample treatment ({\it e.g.} ultra-high vacuum (UHV) conditions).
However, the lack of atomic-level lateral definition in the resist
pattern makes extension of EBL well into the sub-10~nm width regime
exceedingly challenging.

\subsection{Scanned probe lithography - additive}

Lithographic techniques using scanned probes\cite{Quate97SS,SohetAl01}
have also been employed to create sub-100~nm wide nanowires, though no
scanned probe lithography (SPL) approach has yet attained the
wide-spead popularity of EBL.  Scanned probes ({\it e.g.}  atomic
force microscopy (AFM) and scanning tunneling microscopy
(STM)\cite{Bai00}) are natural candidates for adaptation to
fabrication techniques, since the probe tip interacts with the
substrate on a very local scale.  A number of results have been
reported in which AFM and STM have been used to modify a substrate
electrochemically (see Refs.~\cite{AbadaletAl98APA,AvourisetAl98APA}
and references therein).  However, we restrict our discussion to
methods that result in nanowires as defined in
Section~\ref{sec:intro}.

\subsubsection{AFM methods}

AFM-based lithography techniques used for nanowire creation use three
main approaches for pattern definition: direct drawing of metal onto a
substrate; mechanical modification of a resist layer followed by
evaporation and liftoff; or electrochemical modification of a resist
layer followed by evaporation and liftoff.

Using an AFM tip in contact mode to draw various organic molecule
``inks'' directly onto a substrate is a technique known as ``dip-pen
nanolithography'' \cite{PineretAl99Science}.  This method uses the
adsorbed layer of water on the AFM tip under ambient conditions to
``wick'' ink directly to the contact point between tip and substrate.
An approach similar in spirit has been
demonstrated\cite{RamspergeretAl01APL} to produce Au nanowires 4~nm
wide, 1~nm thick with lengths of several microns.  The substrate is
(111) Si that has been heated in vacuum to remove the native oxide and
trigger the $7 \times 7$ surface reconstruction.
At room temperature under UHV, an AFM tip coated by e-beam
evaporation with Au is then drawn across the Si surface in contact
mode.  Gold atoms have sufficient surface diffusivity at room
temperature that a gold wire is left behind on the Si surface where
the tip had been in contact.  While this technique is quite
interesting, it must be performed under UHV conditions, and should
only work with metals having high atomic diffusivities.

\begin{figure}
\begin{center}
\includegraphics[width=8.5cm,clip]{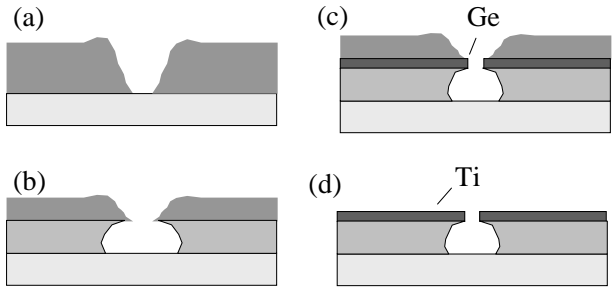}
\end{center}
\caption{Resist profiles associated with different AFM lithography schemes. (a) mechanical deformation, single layer resist.  Note the raised edges around 
the trough, caused by ``plowing''; (b) bilayer resist, as in Ref.~\protect{\cite{SohnetAl95APL}}; (c) plasma-assisted multilayer approach, as in Ref.~\protect{\cite{BouchiatetAl96APL}}; (d) a simpler multilayer method, as in Ref.~\protect{\cite{HuetAl98JVSTB}}.}
\label{fig:spl}
\end{figure}

Mechanical deformation of resist by an AFM tip typically leads to a
resist profile that is not conducive to liftoff processing (see
Fig.~\ref{fig:spl}).  With a trough-shaped indentation, subsequently
evaporated metal tends to cling strongly to the resist sidewalls,
making a clean liftoff problematic.  One approach to dealing with this
issue is that of Ref.~\cite{SohnetAl95APL}.  Rather than ``plowing''
through a single layer of resist, the authors employ a bilayer resist
scheme, using a lower layer composed of PMMA-MAA (methacrylic scid)
copolymer with a considerably higher solubility than the top layer of
PMMA.  The cantilever tip plows through the upper layer, exposing the
more soluble base layer for removal.  The result is an undercut resist
profile that allows good liftoff.  The thick resist bilayer is also
useful for producing wires that cross substrate topography.  The
authors produced 40-50~nm thick and wide nanowires extending over
50~nm-thick predefined contact pads.

Another way to mitigate the resist profile issue is described in
Ref.~\cite{BouchiatetAl96APL}.  A 200-300~nm thick layer of PMMA-MAA
copolymer is spun onto the substrate, baked, and then coated by
thermal evaporation with a 5-15~nm thick germanium layer.  This Ge
layer will eventually act as a suspended mask.  The substrate is then
coated with a 15-20~nm thick polyimide layer and baked at
60$^{\circ}$~C to drive off residual solvent.  This polyimide is
mechanically furrowed by a Si AFM cantilever, exposing the Ge layer
under the polyimide.  By reactive ion etching (RIE), the
exposed Ge is removed, revealing the underlying PMMA-MAA; subsequent
RIE in an O$_{2}$ plasma etches the PMMA-MAA vertically down to the
substrate, and laterally to some extent.  The suspended Ge layer now
acts as a properly undercut mask for metallization and subsequent
liftoff.  Wires as narrow as 40~nm were made with this process.

A slightly simpler approach is described in Ref.~\cite{HuetAl98JVSTB}.
An ultrathin (3~nm) Ti layer is evaporated on top of a baked PMMA
resist layer.  The AFM tip mechanically removes the Ti by abrasion as
the cantilever is drawn across the surface in the desired pattern.
The PMMA exposed by Ti removal is then etched by oxygen RIE to form an
undercut resist pattern.

These mechanical-deformation-based AFM methods succeed as flexible
approaches, capable of producing nanowires substantially below 100~nm
in width on a variety of substrates.    In terms of ultimate wire
size, however, their capabilities appear to differ insignificantly
from those of EBL.  While AFMs may have atomic resolution in
imaging\cite{SugawaraetAl95Science} because of the extremely
short-range nature of the hard-core part of the tip-surface
interaction, no convenient method currently exists to use that
resolution in a patterning mode.  Mechanical surface modification
tends instead to affect a region {\it at least} as large as the radius
of curvature of a typical AFM tip, on the order of 10-20~nm.

Both conducting AFM tips and STM tips have been used as local sources
of electrons to modify a resist layer chemically, in a proximal probe
form of EBL.  Examples of conducting AFM or STM to expose a resist
layer include
Refs.~\cite{MajumdaretAl92APL,KimetAl98JKPS,DavidssonetAl99ME,DuboisetAl99SSE}.
This approach runs into similar limitations as standard EBL, described
in Sec.~\ref{sec:EBL}.  Despite extremely local tip-resist
interactions, the actual wire widths seen in the best
liftoff-processed samples remain near 50~nm.  The need to sink current
from the tip into the substrate at comparatively low voltages means
that this method tends to work best on conducting substrates.
Furthermore, since tip voltages in this method cannot approach those
used in high resolution EBL, achieving a resist profile favorable for
liftoff is challenging.  Similarly, the thinner resist layers
best-suited for this method also tend to be those least suited to
metallization and liftoff.

\subsubsection{STM methods}

One additional scanned probe approach that does seem capable of
producing nanowires with exceedingly narrow cross-sections uses
UHV STM on a conducting silicon substrate.  The patterning
process\cite{LydingetAl94APL} begins with a doped (100) Si wafer in
UHV that has had its native oxide removed by heating to
1250~$^{\circ}$C for 1 minute.  The surface is then exposed at
650~$^{\circ}$C to atomic hydrogen obtained by cracking a dose of
molecular hydrogen  with a
1500~$^{\circ}$C filament.  The hydrogen passivates the silicon
surface, forming a uniform monohydride layer that will act as a
resist.

\begin{figure}
\begin{center}
\includegraphics[width=8.5cm,clip]{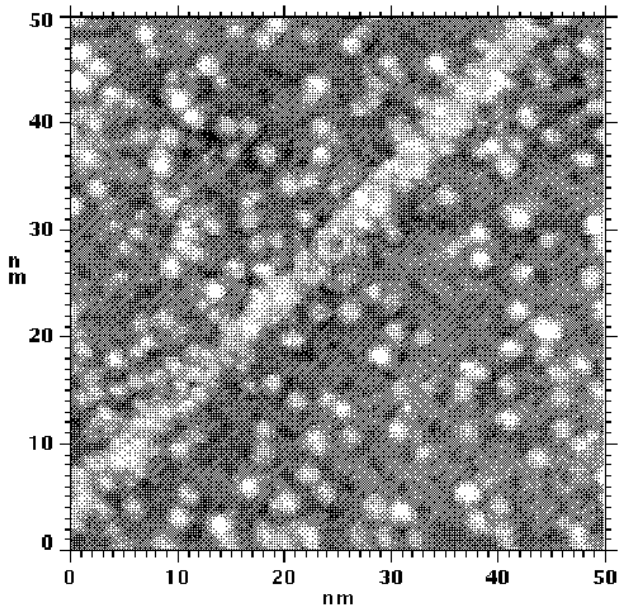}
\end{center}
\caption{An STM image of a cobalt nanowire 3~nm wide on a Si surface,
produced by selective depassivation and surface decoration in UHV.
Figure reproduced with permission from
Ref.~\protect{\cite{PalasantzasetAl99JAP}}, copyright 1999, American
Institute of Physics.}
\label{fig:stmlith}
\end{figure}

The STM tip is then biased negatively with respect to the sample by
several volts\cite{LydingetAl94APL,MasuetAl94JVSTB,AdamsetAl96APL} and
current doses on the order of 100-1000~$\mu$C/cm are used to strip the
hydrogen away through electron-stimulated
desorption\cite{BeckeretAl90PRL,Bolland90PRL}.  The depassivated Si
surface is now available for metallization.  Metal nanostructures have
been produced with this approach using chemical vapor deposition
\cite{MasuetAl94JVSTB,AdamsetAl96APL} as well as physical vapor
deposition\cite{PalasantzasetAl99ME,PalasantzasetAl99JAP}.  In the
former, the chemistry that deposits the metal can only occur on the
depassivated Si sites.  In the latter case Co was deposited at
sub-monolayer coverage, and annealing allowed the Co to migrate to the
unpassivated Si sites and possibly form metallic silicides.  Cobalt
nanowires as narrow as 3~nm were reported, though no transport data
has been presented on such structures; see Fig.~\ref{fig:stmlith}.

The STM depassivation technique is the only SPL
approach thus far that can approach true atomic resolution in
structure definition.  The restrictions on the method, however, are
quite severe: UHV working conditions, a Si substrate that conducts
well enough for STM, and careful control of the metallization process.

\subsection{Scanned probe lithography - subtractive}

STM\cite{SugimuraetAl93JJAP} and conducting
AFM\cite{SnowetAl95Science,WangetAl95APL,SnowetAl96APL,HeldetAl98PE}
have also been employed in {\it subtractive} approaches to nanowire
fabrication.  The resulting devices often do not met our definition of
``nanowire'' due to their small length.  We briefly describe the
method of Ref.~\cite{SnowetAl95Science} here, since that work
specifically addresses devices that satisfy the aspect ratio criterion
of Sec.~\ref{sec:intro}.

A substrate of interest, in this case an oxidized Si wafer, is coated
with a metal layer, 7~nm of Ti, patterned by photolithography.  To
eventually make a narrow wire, one starts with a wider (micron-scale)
structure between larger pads; this allows wire conductance to be
monitored in real time as fabrication procedes.  A conducting AFM tip
is then brought into contact with the metal surface under ambient
conditions (in air, 40\% relative humidity, 300~K).  When the tip is
negatively biased by $\sim$10-12~V with respect to the Ti layer, an
electrochemical reaction\cite{SugimuraetAl93JJAP} takes place due to
the adsorbed water at the tip-metal junction.  The Ti layer
immediately in contact with the tip is converted into TiO$_{2}$, with
a sheet resistance greater than 1~T$\Omega$.  Using the computer to
sweep out particular paths, a Ti film originally 2~$\mu$m wide was
selectively oxidized, leaving an unoxidized wire 15~nm wide, 7~nm
thick, and 500~nm long with a resistance of approximately
100~k$\Omega$.  Wires with nominal widths as narrow as 3~nm are
reported in this work.  Variations of this approach have been used on
chromium\cite{WangetAl95APL} and aluminum\cite{SnowetAl96APL}.

As pointed out in Ref.~\cite{SnowetAl95Science}, inferring wire
transverse dimensions produced in this technique from resistance
measurements is nontrivial.  In the smallest structures actual
conducting cross-section does not scale linearly with apparent wire
width.  Reasons for this include the intrinsic roughness of the wire
edges due to granularity of the metal, and swelling of the anodized
material because of lattice mismatch between the metal and its oxide.
In fact, metallic conduction is observed even in structures so narrow
that oxide swelling has eliminated any AFM-image signature of
unoxidized material.  Feedback control of anodization using
the {\it in-situ} measured wire conductance allows the electrical
properties of resulting devices to be precisely controlled, but
geometrical measurements of the wires are not readily performed 
when widths fall below 10~nm.

\subsection{Lithographic methods:  summary}

Some general observations are possible.  First,
the prime advantage of lithographic approaches 
is the flexibility in nanowire geometry that is then possible.  By
definition lithographic methods are capable of producing nanowires of
nearly any shape that may be ``drawn''.  Second, resist-based
techniques face fundamental limitations due to the development process
and the need for a resist profile that will generate clean liftoff.
Third, while probes like AFM and STM are capable of extremely
local examination of the properties of substrates, employing these
generalized tools for precise lateral definition of nanowire
structures is nontrivial.  SPL often requires special substrates,
metals, UHV conditions, or some combination of the three.  Finally,
even with atomically precise lateral definition of a hole in resist
(such as STM-depassivated Si), the ultimate limits of nanowire size
are set by the surface science of the constituent metal itself, and of
the metal-substrate interface.

\section{\small{TEMPLATING}}
\label{sec:template}

Templating has also proven extremely useful for the fabrication of
nanowires.  While a lithographic technique employs some kind of
drawing to define the pattern, templating instead utilizes nanoscale
surface relief to provide definition to the metal structure.  The keys
to nanoscale wire construction then become finding a suitable template
and controlling metal morphology.  Below we discuss a number of
templating approaches.

\subsection{Step-edge lithography}

\begin{figure}[h]
\begin{center}
\includegraphics[width=8.5cm, clip]{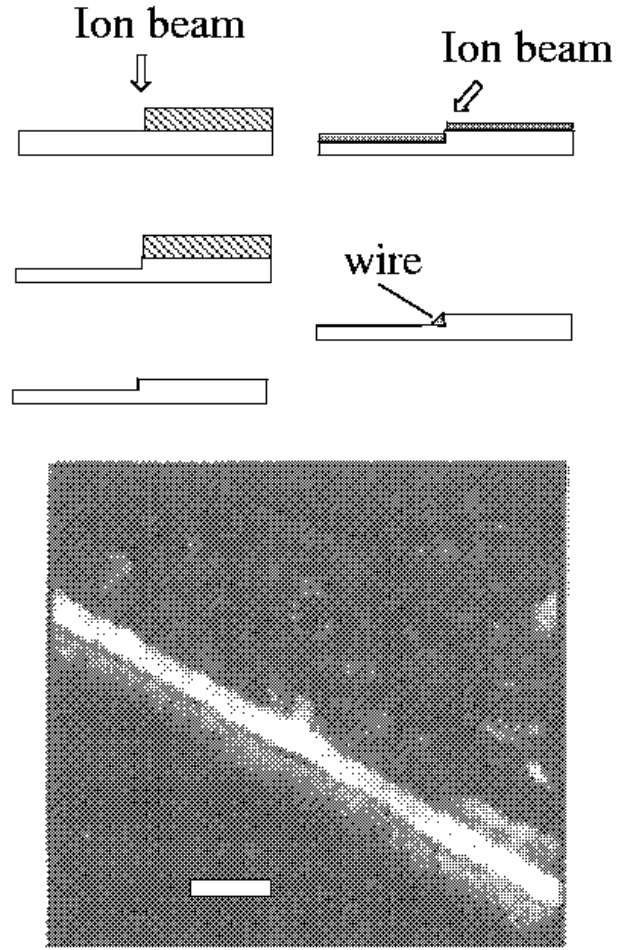}
\end{center}
\caption{Step-edge lithography templated fabrication method.  The
image shows a resulting AuPd wire (scale bar = 100~nm).  Figure
reproduced with permission from \protect{\cite{Giordano80PRB}},
copyright 1980 American Physical Society.}
\label{fig:stepedge}
\end{figure}

One templating approach that has proven quite successful and
innovative is ``step-edge lithography''\cite{ProberetAl80APL} (SEL),
as illustrated in Fig.~\ref{fig:stepedge}.  On a suitable substrate
such as a glass slide, an initial patterning step is used to define an
edge; for example, photolithography and liftoff processing may be used
to define a large chromium pad on the substrate.  The sample is now
placed in an ion mill, and a beam of Ar ions is directed to impinge at
normal incidence on the substrate.  Because of the differential etch
rate between the substrate and the chromium, the area covered by the
chromium becomes a mesa, while the exposed substrate surface is
sputtered away by the ion beam.  The chromium is then removed with a
standard wet etch.  This has now produced surface relief on the
substrate (a crisp step at the edge of the mesa) with a critical
dimension (the step height) determined by the ion etch time.

Next, metal is deposited at normal incidence onto the substrate.  As
with all such metallization steps, to achieve transverse wire
dimensions on the few-nm scale requires the metal to have nm-scale
grains or be amorphous.  A AuPd alloy was first used to demonstrate
this technique.

Now the metal-coated substrate is placed back in the ion mill, with
the ion beam this time incident on the sample at a substantial angle
($\sim 45^{\circ}$) from the normal.  The ion mill then sputters away
the exposed metal on top of the mesa as well as that on the remainder
of the substrate, with the exception of one place: The step at the
mesa edge geometrically protects the metal right next to the step from
the etching effects of the ion beam.  The ideal end result is a
nanowire of triangular cross-section running the length of the mesa
edge.  Again, the critical wire dimensions have been determined by
metal deposition (thickness) and template geometry (width via the etch
angle).  Contacts to larger pads may then be made via additional
lithography steps or other means.

This method has been employed to make 10~nm-scale nanowires from a
variety of materials, including normal
metals\cite{GiordanoetAl79PRL,FlandersetAl81JVST,WhiteetAl82PRL}, superconductors\cite{GiordanoetAl89PRL},
and ferromagnetic metals\cite{HongetAl95JMMM}.  Nanowires made with the
step-edge approach may be hundreds of microns long.  A limitation is
that, because of the shadowing involved, the lateral definition of the
wire is only as good as the collimation of the incident ion beam.
Similarly, by necessity the ion beam damages the exposed wire surface.
Furthermore, the edge roughness of the original step is set by the
initial lithography and chrome metallization processes.  Finally, this
technique works best if the ion etch rate of the deposited metal is on
the order of or faster than that of the substrate.

A clear advantage of this technique is its ability to reach extremely
small lateral length scales while preserving a large aspect ratio.
The tradeoff when compared to EBL or SPL is one of flexibility;
step-edge lithography is very good for achieving linear nanowires in
isolation, rather than for producing multiple nanowires intersecting
at arbitrary angles.

\subsection{MBE-defined templates}
\label{sec:mbe}

\begin{figure}[h]
\begin{center}
\includegraphics[width=8.5cm,clip]{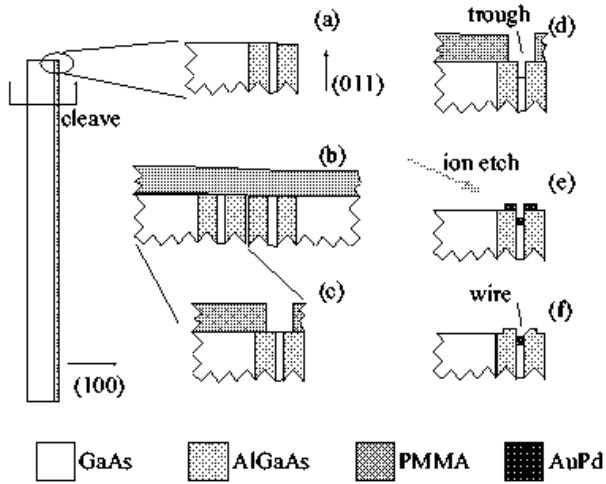}
\end{center}
\caption{Schematic showing the steps in the MBE-template fabrication
process.  An MBE-grown substrate is cleaved; lithography on the
cleaved surface is used to expose the thin GaAs layer, which is
selectively wet-etched to produce a trough; wire material is
deposited, the resist is lifted off, and a directional ion etch
removes excess material; a nanowire is left in the trough, ready for
further lithography to define multiple leads.}
\label{fig:cartoon}
\end{figure}

A recent innovation in nanowire fabrication with a number of
similarities to the step-edge technique is described in
Refs.~\cite{NatelsonetAl00SSC,NatelsonetAl00APL}.  However, rather
than relying on lithographic and ion etch processes to provide lateral
definition to the nanowires, the authors take advantage of the
remarkable thickness resolution possible in molecular beam epitaxy
(MBE).  In the GaAs/AlGaAs system monolayer thickness control of MBE
growth is typical.  This MBE templating approach is shown in
Fig.~\ref{fig:cartoon}.

This technique begins with a 500~$\mu$m thick undoped (100) GaAs
substrate, onto which has been grown a layered structure.  To produce
single wires of width $d$, the layers are: 2~$\mu$m thick
Al$_{0.3}$Ga$_{0.7}$As, a GaAs layer of thickness $d$, and an
additional 2~$\mu$m of Al$_{0.3}$Ga$_{0.7}$As.  All of these layers
are undoped.  The substrate is then cleaved into a strip approximately
5~mm by 15~mm to expose the (0$\overline{1}$1) face.  The growth layer
thickness $d$ is now the width of a GaAs strip on the
(0$\overline{1}$1) surface.

The strip is again cleaved to produce two 5~mm by 7.5~mm pieces, which
are then secured to each other by epoxy with their (100) faces
touching.  This protects the critical edges of the pieces while
leading to a combined area of $\sim$~1~mm by 7.5~mm of
(0$\overline{1}$1) surface on which to work.  The (0$\overline{1}$1)
faces are spin-coated with 4.5~\% 950K PMMA in a $\sim$~1~$\mu$m thick
layer, to provide uniform coverage.  EBL is used to expose the
strip of width $d$ on the (0$\overline{1}$1) surface. 

The assembly is then etched at room temperature in a solution of
500~mL 30\% H$_{2}$O$_{2}$ / 400~$\mu$L 30\% NH$_{4}$OH.  This
solution etches bulk GaAs at a rate of $\sim$ 45~nm/min, while etching
Al$_{0.3}$Ga$_{0.7}$As approximately 100 times more slowly.  Following
this etch what had been a GaAs strip on the (0$\overline{1}$1) surface
is now a trough with a width determined by MBE growth and a depth set
by the etch time.  A depth-to-width ratio of approximately 2.5:1 has
been found to be optimal for the remainder of the fabrication process.
This trough will act as a mechanical template, providing lateral
definition for the nanowire.  TEM examination\cite{Natelsonunpub} has
shown that the trough walls and floor are lined with a $\sim$ 1~nm
thick oxide layer left behind by the wet etch.

Metal is then deposited by e-beam evaporation or sputtering at normal
incidence to the (0$\overline{1}$1) surface, into the trough and onto
the surrounding exposed wafer material.  After liftoff of the PMMA
layer, the assembly is placed in an ion etching system (either an RIE
or an ion beam) so that the ions are incident at approximately
60$^{\circ}$ from normal to the (0$\overline{1}$1) plane.  The
surface geometry protects the metal in the bottom of the trough, while
the excess metal on the (0$\overline{1}$1) surface is sputtered away.
Additional EBL and liftoff processing may be used to form leads to
monitor wire conductance, or to fabricate nearby gates.  At the
conclusion of the directional dry etching process, the result is a
nanowire in a trough.

\begin{figure}
\begin{center}
\includegraphics[width=8.5cm,clip]{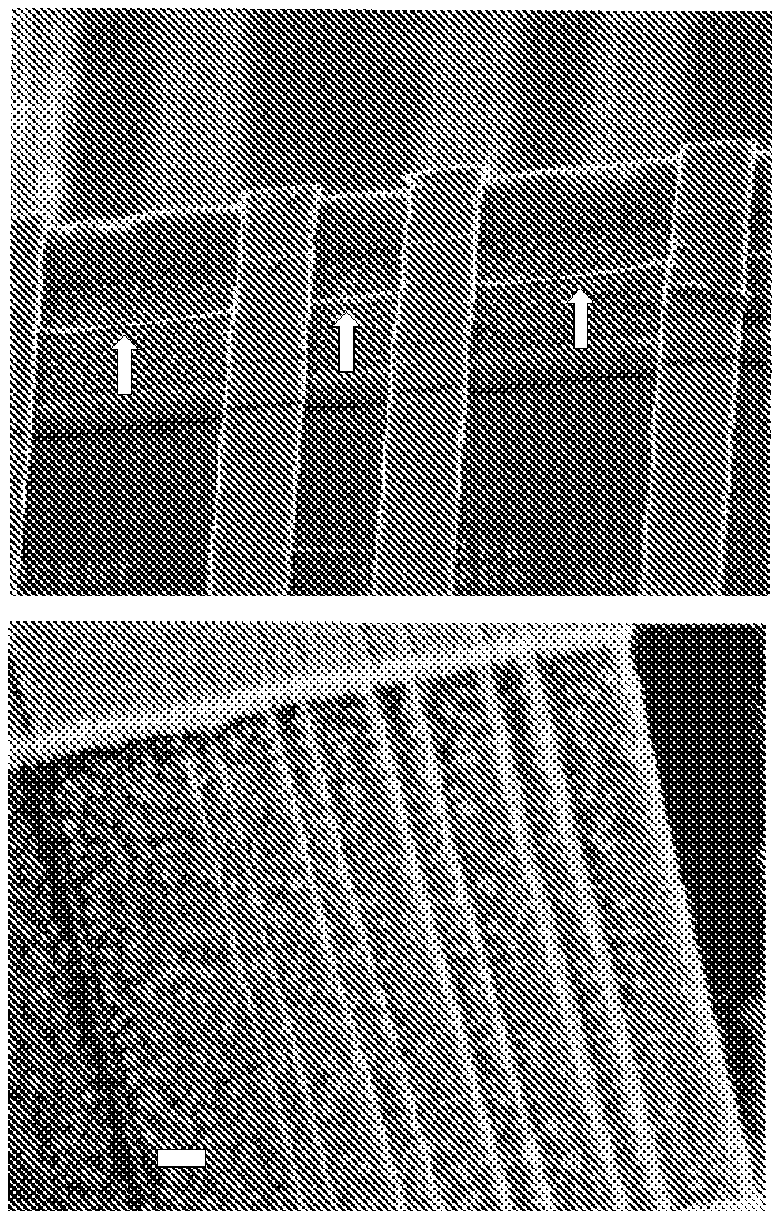}
\end{center}
\caption{Upper image: a 20~nm wide AuPd wire (indicated by arrows) made with the MBE templating process, spanning several EBL-fabricated Au leads.  Surface etching of the GaAs/AlGaAs substrate has been performed to aid in imaging.  Lower image:  an array of parallel nanowires made with the same technique.  The largest wires are 50~nm wide, while the narrowest are 7.5~nm wide.  Scale bar = 100~nm.}
\label{fig:mywire}
\end{figure}

A different wet etch may be used to remove some of the surrounding
Al$_{0.3}$Ga$_{0.7}$As, allowing SEM characterization of the nanowire;
see Fig.~\ref{fig:mywire}.  With more complicated layered structures,
arrays of nanowires may be fabricated in parallel, shown in
the lower portion of Fig.~\ref{fig:mywire}.

This technique requires specialized substrates, and, like step-edge
lithography, is clearly best suited for producing single or arrays of
parallel wires.  One key difference between this approach and SEL is
that here the wire width is defined by the MBE growth process, rather
than the collimation of the directional etch.  As a result, the wire
cross-section tends to be rectangular rather than triangular, with the
ultimate limit of accessible wire size set by the morphology of the
deposited metal.  Using e-beam evaporated Au$_{0.6}$Pd$_{0.4}$ with a
1~nm Ti adhesion layer, wires as small as 5~nm wide and 7.5~nm thick
with lengths well over 1~$\mu$m have been studied extensively with
multiterminal transport measurements\cite{NatelsonetAl01PRL}.
Some continuous two-terminal devices as narrow as 3~nm have been 
made, though they were extremely fragile\cite{Natelsonunpub}.
MBE-grown templates combined with either amorphous or epitaxially
deposited metals have the potential to create engineered nanowire
structures as narrow as a few atoms.

\subsection{Nanoscale etch-masks}

Another subtractive approach to templating 
is to employ nanoscale etch masks.  The idea is to coat an entire
substrate with a thin metal layer.  One deposits (by spinning
from solution) some nanowire-shaped templates onto the surface.
The substrate is then placed into a dry etcher, either an ion mill
or a RIE, and the templates protect nanowire-shaped patches of
the surface from the normally-incident beam.  Finally, the templates are
removed, ideally leaving behind nanowires where the underlying 
metal film had been protected.

Multiwalled carbon nanotubes have been used as templates for such a
process\cite{YunetAl00JVSTA}.  Beginning with a (111) Si wafer coated
with 200~nm of thermal oxide, a 1~nm Ti adhesion layer and a 9~nm Au
film were evaporated onto the substrate.  A few drops of multiwalled
nanotubes suspended in CHCl$_{3}$ were spin-coated onto the surface,
leaving behind isolated nanotubes as the solvent evaporated.  The
substrate was then placed in the path of a 300~V beam of Ar$^{+}$ ions
at normal incidence for 1 minute, long enough to mill away the metal
film and $\sim$~1~nm of SiO$_{2}$ from uncovered portions of the
surface.  The nanotubes of various diameters act as etch masks,
allowing isolated Ti/Au wires as narrow as 12~nm to be produced.  The
remaining nanotubes could be removed by AFM manipulation, though one
suspects that exposure to an oxygen plasma or ozone environment might
be more effective at speedily ridding large areas of any carbon
residue.

A variation on this approach\cite{SordanetAl01APL} addresses two
difficulties in the above technique: the dispersity of multiwalled
nanotube sizes and the difficulty of removing carbon residue.  Rather
than carbon nanotubes, the templates are chemically modified
V$_{2}$O$_{5}$ fibers.  Unlike fibers produced in pure water, these
are hydrolyzed in a dilute solution of {\it N}-methylforamide,
resulting in transverse fiber dimensions of 6-10~nm $\times$ 15-20~nm,
with lengths on the micron scale.  These authors use a 1~nm Ti / 6~nm
AuPd film for enhanced metal uniformity from the small AuPd grain
size, and an Ar ion beam (200~V) to provide the directional etch.  The
V$_{2}$O$_{5}$ fibers consistently lead to fairly monodisperse
15~nm-wide nanowires whereever the fibers protect the film from the
etch.  Residual V$_{2}$O$_{5}$ is removed by soaking the substrate in
a dilute acid solution.

A more involved example of this approach\cite{ChoietAl00JVSTA}
interposes an additional step.  Starting with a CdTe substrate coated
with a Bi film (the desired wire material; the unusual substrate
facilitates MBE growth of high quality Bi films), a layer of PMMA is
spun on and baked.  The masking in this case is provided by chemically
self-assembled chains of Ag nanocrystals.  The chains are $\sim 40$~nm
in diameter, with micron-scale lengths.  These
structures\cite{HeathetAl97JPCB} are transferred from solution as a
Langmuir-Schaeffer film onto the PMMA-coated substrates that are then
flood-exposed with a dose of 50~$\mu$C/cm$^{2}$ of 700~V electrons.
This is sufficiently energetic to expose the PMMA but not to penetrate
the Ag nanocrystals.  The nanocrystals act as a shadow mask for e-beam
exposure rahter than as a direct etch mask.  After the exposed resist
is removed by development in MIBK:IPA solution, the substrate is
etched with BCl$_{3}$ in a RIE etcher.  The BCL$_{3}$ attacks the
exposed Bi, leaving Bi nanowires behind.  Removal of the remaining
PMMA and Ag nanocrystals is not discussed.

A final example of this approach\cite{FritzscheetAl99APL} uses
biopolymeric materials as etch masks.  Microtubules with diameters
smaller than 40~nm and lengths of tens of microns were used to shadow
regions of Ti/Au film on an oxidized Si wafer.  Biopolymer templates
open up the possibility of using biologically inspired pattern
formation mechanisms to engineer complicated mask structures.

The advantage of approaches like these is that the templates may be
fabricated by chemical means, often with high reliability and
uniformity on the nm scale.  The fidelity with which this uniformity
is transferred to the final structures is then limited by the
etching technique and the morphology of the initial metal film.
A further complication is the positioning of the templates; the
methods described here form wires whereever the templates happen
to land, rather than in predetermined locations.

\subsection{Suspended nanotubes}

One may also consider templating approaches that are additive rather
than subtractive.  Here a template is used as a mechanical scaffold
upon which wire material is deposited.  Using suspended carbon
nanotubes in this way has allowed the creation of amorphous MoGe
nanowires with widths from a few nm up to
$\sim$~20~nm\cite{BezryadinetAl00Nature}.

For this technique to produce nanowires suited for conductance
studies, the nanotube must be suspended over the substrate so that it
is electrically isolated, except at end points.  One method of
achieving this configuration\cite{BezryadinetAl97APL} begins with a Si
wafer coated with 1~$\mu$m of SiO$_{2}$ and a further 60~nm of
Si$_{3}$N$_{4}$ film.  EBL and RIE are used to open a $\sim$100~nm
wide slit in the nitride with a narrower constriction where the
measuring electrodes will be.  An HF treatment etches the exposed
SiO$_{2}$ and undercuts the Si$_{3}$N$_{4}$.  The result is a trough
with two cantilevered Si$_{3}$N$_{4}$ protrusions at the location of
the original constriction.  Through a shadow mask, metal is deposited
to coat these protrusions, forming two measuring electrodes spaced
very closely, with the HF etch undercut preventing the two electrodes
from shorting together.  Electrostatic trapping is then used to
``capture'' a single nanotube from solution to bridge the 
electrodes.

\begin{figure}[h]
\begin{center}
\includegraphics[width=8.5cm,clip]{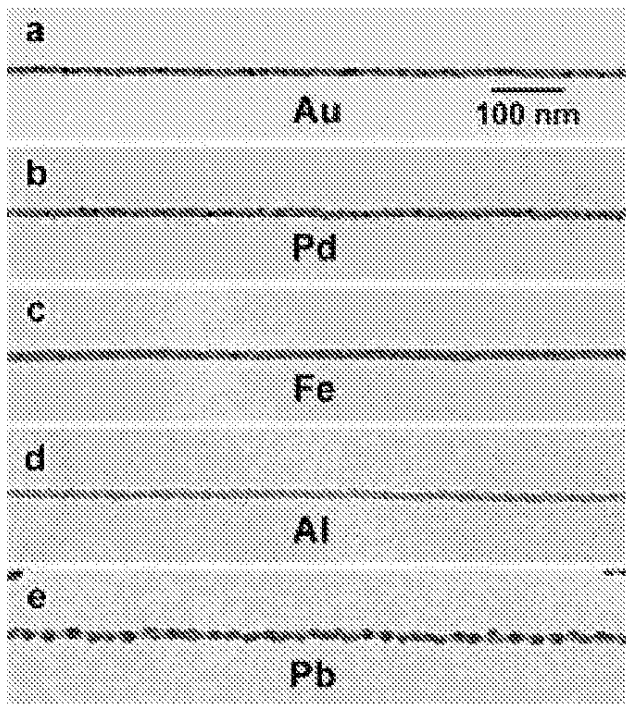}
\end{center}
\caption{Nanowires made using suspended single-walled carbon nanotubes
as templates for deposition.  Tubes were coated with 2~nm Ti and 5~nm
of the designated metal.  Figure reproduced with permission from
\protect{\cite{ZhangetAl00APL}}.  Copyright 2000, American Institute
of Physics.}
\label{fig:susptube}
\end{figure}

An alternative approach\cite{ZhangetAl00APL} uses chemical vapor
deposition to grow nanotubes directly from patterned catalyst-coated
islands on a Si wafer\cite{KongetAl98Nature}.  The direct growth
method is easier to use for longer nanotubes than electrostatic
trapping.

As we have seen in all other techniques, in the limit of an atomically
precise template the quality of narrow nanowires can be limited by the
morphology of the constituent metal.  That is again true for this
approach.  Zhang {\it et al.}\cite{ZhangetAl00APL,ZhangetAl00CPL} have
studied the metal-nanotube interface by preparing a series of
nanowires from Au, Pd, Fe, Al, and Pb both with and without a 1-2~nm
Ti buffer layer.  Metals deposited on top of the Ti adhesion layer are
much more uniform, leading to continuous nanowires at coverages of
$\sim$5~nm of the working metal (see Fig.~\ref{fig:susptube}).  They
postulate that the electronic structure of Ti makes it particularly
well-suited to forming thin, uniform coatings on the graphite-like
surface of the nanotubes.

Suspended templates such as these are versatile and enable the
formation of extremely narrow structures.  With electrostatic trapping
or patterned catalyst deposition, the nanowire position may be
engineered rather than determined by chance.  UHV conditions are not
necessary for this approach.  However, the need to electrically
isolate the templates from the substrates limits the geometric
flexibility of the technique.  Measurements with more than two
electrodes in contact with the wire are also challenging 
in this configuration.

\subsection{Electrodeposition into channels}

  Electrochemical approaches have also been used very effectively to
produce nanowires using additive templating.  To constrain the growth
of electrochemically deposited metal, a three-dimensional template
({\it i.e.} a channel with transverse dimensions close to the desired
final wire size) is most commonly required. 

  An ideal template would be a long pore, ideally of uniform
cross-section, through an electrochemically inert membrane.  A recent
review of membrane-based templating of nanoscale materials may be
found in Ref.~\cite{Martin94Science}.  One
approach\cite{Possin71Physica}, borrowing from a nuclear physics
technique of particle detection, is to create such pores by chemically
etching membranes that have been exposed to an energetic ion flux from
a radioactive source\cite{FleischeretAl75}.  The MeV-scale ions pass
through thin films of mica or polycarbonate, leaving behind a
chemically-altered track.  Mica tracks are etched by HF, while
polycarbonate membranes may be etched by a mixture of NaOH and methyl
alcohol.  Pore sizes below 100~nm are routine, and diameters below
10~nm have been reported\cite{WilliamsetAl84RSI}.  Polycarbonate
membranes may also be purchased commercially ({\it e.g.} Nuclepore
Corp. or Poretics Corp.).  Pore density is determined by the
integrated ion dose and may be as high as $10^{9}$~cm$^{-2}$, while
diameter uniformity is set by the details of the etch
process\cite{SchonenbergeretAl97JPCB}.

\begin{figure}
\begin{center}
\includegraphics[width=8.5cm,clip]{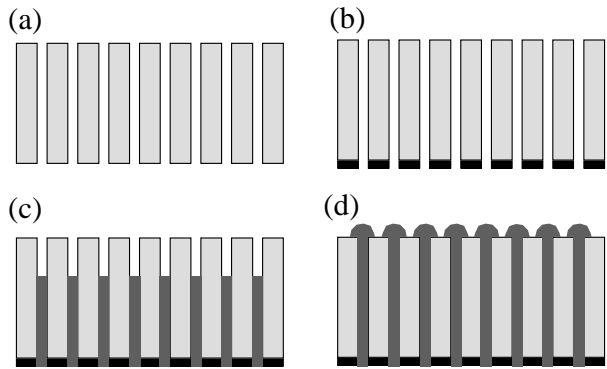}
\end{center}
\caption{Nanowires by electrodeposition into a porous template.  (a) a nanoporous membrane; (b) a seed layer of metal sputtered or evaporated onto one
side of the membrane; (c) electrodeposition in progress - metal composition
may be varied by appropriately changing the deposition solution; (d) completed
nanowires still in template.}
\label{fig:channel}
\end{figure}

  An alternative membrane material is anodically etched porous
alumina.  This may be produced electrochemically from a thin aluminum
membrane\cite{Despic89}; such membranes may purchased commercially,
albeit with a limited range of pore sizes, from, {\it e.g.}, Whatman
Lab. Division.  Pore densities can be as high as $10^{11}$~cm$^{-2}$,
with diameters as small as a few nm\cite{Martin94Science}, and
alumina's chemical stability allows great diversity in the types of
materials that may be deposited\cite{RoutkevitchetAl96IEEE}.  Another
template alternative is nanochannel
glass\cite{TonuccietAl92Science,NguyenetAl98JECS}, in which pores are
reduced from microscopic to nanoscale dimensions by drawing of the
glass template.

   Chemical self-assembly has also been employed to create templates
for electrodeposition.  Diblock copolymers\cite{FasolkaetAl01ARMR} are
composed of two polymer components that, under appropriate temperature
and surface conditions, phase separate into self-organized regions
containing the different polymers.  Because of the chemical
distinctiveness of the regions, with suitable processing
such self-assembled structures may be used to create templates for
nanowire deposition.  For example\cite{ThurnAlbrechtetAl00Science}, a
diblock copolymer composed of PMMA and polystyrene (PS) dissolved in
toluene may be spin-coated onto a conducting substrate.  By
application of an electric field normal to the substrate surface, the
self-organization produces a 0.5~$\mu$m-thick, hexagonally close-packed
array of PMMA rods 14~nm in diameter embedded in a PS matrix.
Exposure to DUV radiation degrades the PMMA and crosslinks the PS, so
that subsequent development in acetic acid leaves behind a dense array
of pores in a PS film for use as a template.  Pore density and
parallelism is similar to that in alumina films, while not requiring
strong acids or bases for processing.
  
   Figure \ref{fig:channel} illustrates the steps in the templating
process.  An initial metal layer is deposited by evaporation or
sputtering onto one side of the membrane.  The membrane is then placed
in an appropriate solution and the desired metal is electrochemically
deposited, using the initial metallization as a seed.  Note that wires
with longitudinally varying chemical composition may be grown by
changing the deposition solution during the growth process.  Precise
multilayers are possible and may be combined with chemical
functionalization for self-assembly experiments\cite{MartinetAl99AM}
or diagnostic sensing\cite{NicewarneretAl01Science}.  Similarly, by
depositing alternating layers of ferromagnetic and normal metals, wires
that exhibit longitudinal giant magnetoresistance (GMR) have been
made\cite{PirauxetAl94APL,LiuetAl95PRB}.  It should also be noted that
templates such as these may be filled by methods other than
electrodeposition, such as high pressure injection of molten metal
into the pores (see \cite{ZhangetAl98APL} and references therein).

  Following deposition the membrane may be dissolved with an
appropriate solvent, allowing access to individual nanowires.
Alternately, for sparse arrays made using low-porosity membranes,
lithography on the upper membrane surface may allow two-terminal
contacting of small numbers of
nanowires\cite{WilliamsetAl84RSI,BachtoldetAl98ME}.

  Attractive features of this templating approach include: the ability
to make large numbers of wires in parallel; the ability for those
wires to be substantially narrower than 100~nm with a small dispersity
in wire diameter; and applicability to a large number of material
systems, including noble metals, ferromagnetic materials, semimetals,
etc.  Wire diameter is limited by the template, particularly the
longitudinal uniformity of pore size\cite{SchonenbergeretAl97JPCB},
and the morphology of the electrodeposited material.  It is possible
to form single crystal nanowires with appropriate materials and
conditions; for example, see Refs.~\cite{YietAl99APL,GaoetAl01APA}.
A complication of this method is that arranging individual wires 
in precise configurations for study is quite challenging.

\subsection{Other templated deposition}

  Cleaved heterostructures can be another example of an unconventional
template.  A heavily doped well layer in an MBE-grown III-V
heterostructure on the edge of a cleaved can be used as the active
electrode for deposition wafer\cite{FasoletAl97APL}.  While the
quantum well itself is defined with atomic precision, as discussed
above in Sec.~\ref{sec:mbe}, deposition in this approach is not
capable of producing wires with comparable definition.  The growth of
deposited metal is not constrained laterally, so that the resulting
nanowires produced from a 4~nm-wide quantum well are granular and have
widths $\sim$20~nm.

  Chemically synthesized template structures may also be employed as
scaffolds for nanowire deposition.  Hong {\it et
al.}\cite{HongetAl01SCience} synthesize arrays of 
organic nanotubes from calix[4]hydroxyquinones (CHQs).  The pores in
these arrays are 0.6~nm~$\times$~0.6~nm and are lined with OH groups
and $\pi$-conjugated faces.  The hydrophilic OH groups allow silver
ions in aqueous solution to be intercalated into the pores and reduced
out of solution, where they aggregate to form 0.4~nm diameter
crystalline Ag nanowires up to microns in length.  This approach
produces large numbers of extremely narrow, straight nanowires without
the need for UHV or high temperature processing.  Forming metallic
contacts to the resulting structures, however, is extremely
challenging\cite{Kimpriv01}.
  
Individual molecules may also be used as templates for chemical
metallization.  For example, DNA molecules may be functionalized to
form connections between lithographically defined Au
electrodes\cite{BraunetAl98Nature}.  A chemical procedure exploiting
ion exchange is used to seed silver clusters along the DNA.  Chemical
reduction of additional Ag from solution then leads to the formation
of a granular wire spanning the electrodes.  While this method can
take advantage of existing tools for manipulating DNA, the quality
of the resulting wires is not well-controlled at present.

\section{\small{CHEMISTRY AND SELF-ASSEMBLY}}

Direct chemical synthesis and other self-assembly techniques have also
demonstrated nanowire formation.  We have already discussed templated
chemical synthesis, which relies on a template to confine the
reactants and shape the metallic reaction products.  Direct chemical
synthesis instead takes advantage of some inherent anisotropy in
reaction materials or kinetics to produce extended structures.  Other
forms of self-assembly utilize anisotropic energetics ({\it e.g.}
strain energy from lattice mismatch) to guide nanowire formation.

\subsection{Chemical approaches}

Direct chemical synthesis of nanowires can be difficult to
generalize to large numbers of metals because of the need for precise
control of anisotropy in growth, and a lack of suitable reaction
chemistries.  A full explanation of chemical approaches  
is beyond the scope of this article.  Instead
we very briefly state some essential features of the methods
and provide some references in this rapidly developing field.

Synthesizing elongated structures in either the solution (for example,
\cite{PuntesetAl01Science}) or the gas phase\cite{HuetAl99ACR} often
involves clever use of seed or catalytic particles.  Particle size can
directly influence the diameter of the resulting nanowire.  One
example of catalytic growth that has been very successfully applied to
semiconductor materials (though not yet to metals) is described in
Ref.~\cite{HuetAl99ACR}.  The growth mechanism is called the
vapor-liquid-solid (VLS) technique\cite{WagneretAl64APL,Wagner70}.  A
catalytic liquid nanocluster is made from a material that can form a
liquid alloy with the desired nanowire material.  Such a cluster,
which defines the diameter of the nanowire, is then placed in an
environment supersaturated with reactant, and serves as a nucleation
site for crystallization.  One-dimensional growth then occurs, with
the liquid nanocluster serving as one end of the resulting nanowire.

An alternative method of chemically producing elongated structures
from nanoparticles\cite{PuntesetAl01Science} uses surfactants to coat
the seed.  By using multiple surfactants, the crystallographic
direction of growth of crystalline material from solution onto the
seed may be controlled.  This method has been used to produce single
crystal Co nanorods with diameters below 15~nm, though to date their
lengths have been limited to $\sim$~100~nm.

Other chemical means may be used to encourage growth anisotropy.  For
example, by reducing silver out of solution onto 4~nm seed particles
in the presence of a micellar template\cite{JanaetAl01CC},
microns-long Ag nanowires $\sim$12~nm in diameter may be formed.  

Some metals possess crystallographic structures
that naturally tend toward anisotropic growth when synthesized from
solution.  Examples include selenium\cite{GatesetAl00JACS} and alloys
of selenium and tellurium\cite{MayersetAl01AM}.  These substances tend
to have helical crystal structures that favor 1d growth.  Single
crystal selenium nanowires 10~nm in diameter microns in length have
been produced from solution.

Chemical methods sometimes succeed in producing nanowire structures
even when there is no clear mechanism for growth anisotropy.  Silver nanowires produced from
solution with AgBr seeds\cite{LiuetAl01AM} and zinc
``nanobelts''\cite{WangetAl01CC} from ZnS and graphite powder heated
in flowing Ar are two examples.  

\begin{figure}[h]
\begin{center}
\includegraphics[width=8.5cm,clip]{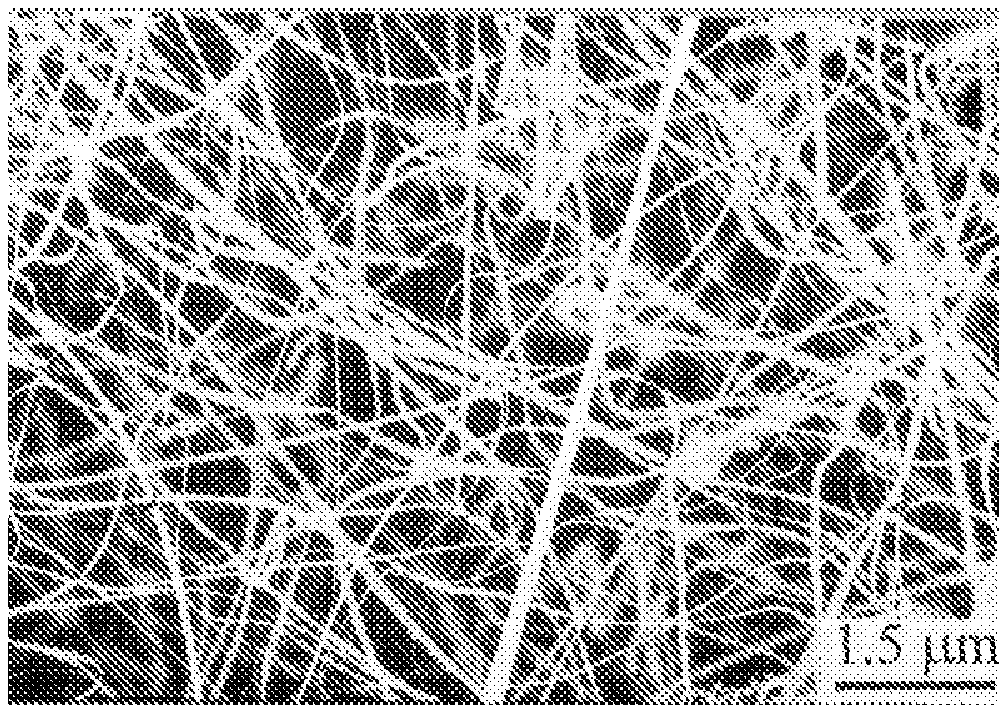}
\end{center}
\caption{SEM image of chemically synthesized Zn ``nanobelts''.
The growth mechanism that produces these anisotropic nanowire structures is
not yet known.  Figure reproduced with permission from
\protect{\cite{WangetAl01CC}}, copyright 2001 Royal Society of Chemistry. }
\label{fig:wang}
\end{figure}

Development of chemical methods for nanowire synthesis are 
in an early stage, and have already produced some spectacular 
results.  As with templating approaches, for study of electrical
transport properties, the nanowires produced chemically must then
be separated and manipulated onto leads.  If suitable reaction
pathways and growth mechanisms may be found and utilized for 
more materials of interest, chemical synthesis may play
a major role in future nanowire studies. 

\subsection{Self-assembly}

  Finally, self-assembly techniques may be employed, particularly to
form ordered arrays of nanowires from certain materials on specific
substrates.  One might imagine that the large configurational entropy
associated with placing adsorbate atoms on a substrate would make
spontaneous formation of nanowire structures unlikely.  Energetic
considerations can make organized patterns favorable, however.  This
section only touches on the richness of this subject, focusing on
recent examples exlicitly dealing with metal nanowires.

  One common means of ensuring linear structures is ``step-edge
decoration.''  Here one considers a substrate that is a vicinal
surface, where the surface normal deviates very slightly from a high
symmetry crystalline direction.  As a result, for clean surfaces
(usually annealed at high temperatures in UHV), the substrate has a
series of parallel atomic terrace steps that are roughly linear, with
the density of steps increasing with the degree of ``mis-cut''.
Nanowire material is then deposited at sub-monolayer coverages.
With appropriate choices of substrate, nanowire material, and
deposition/annealing conditions, deposited metal migrates via 
surface diffusion until being trapped up against step edges.

  Step edge growth has been employed by a number of investigators (for
example,
Refs.~\cite{JungetAl95APA,HimpseletAl97SRL,BatzilletAl98N,DekosteretAl99APL})
in recent years.  The resulting wires are often difficult to
characterize electronically.  The wires are usually only one or two
atomic layers thick perpendicular to the substrate; further, when
metal wires are formed at steps on the surface of another {\it metal},
isolating the conducting properties of the wires from the substrate
may not be possible.

  One interesting variation that evades these difficulties is that
of Zach {\it et al.}\cite{ZachetAl00Science}.  Using electrochemistry
the authors deposit molybdenum oxide at step edges on a cleaved
graphite substrate.  By exposing the substrate to hydrogen at 
500$^{\circ}$~C for an hour, the oxide is reduced to form metallic
Mo wires.  The wire adhesion to the graphite is sufficiently poor
that the wires may be transferred to the surface of a polymer film
cast on top of the graphite substrate.

\begin{figure}
\begin{center}
\includegraphics[width=8.5cm,clip]{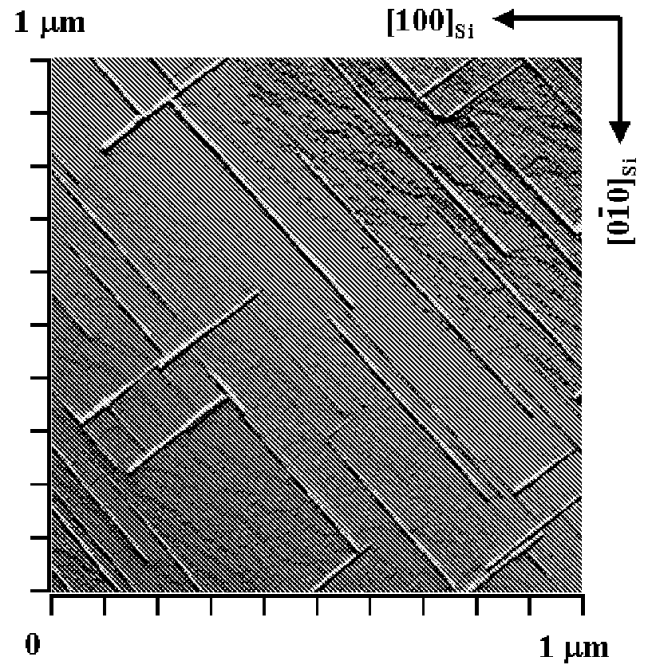}
\end{center}
\caption{STM image of self-assembled ErSi$_{2}$ nanowires on the Si
(001) surface.  The nanowires are less than 5~nm wide and are microns
in length.  Figure reproduced with permission from
\protect{\cite{ChenetAl01MSEB}}, copyright 2001 Elsevier Science. }
\label{fig:silicide}
\end{figure}

  An alternative approach to nanowire fabrication that has received
much recent attention is self-assembly assisted by differential strain
between a deposited metal and the underlying substrate lattice.  As in
the step decoration method, energetic considerations in certain
material systems may be sufficient to overwhelm the entropy gain that
would result from non-straight wires.  Lattice mismatch strain has
been employed extensively in epitaxially grown semiconductor
structures in recent years (see Ref.~\cite{PetroffetAl01PT}).

  Simple lattice mismatch is not sufficient to produce wire
structures\cite{ChenetAl00APL,PetroffetAl94SLM}; isotropic lattice
mismatch results in islands rather than nanowire formation.
Substantial growth anisotropy requires the wire material to have
minimal lattice mismatch along the wire direction and large mismatch
along the transverse direction.  This self-assembly technique requires
epitaxially smooth and clean substrate surfaces, UHV growth
conditions, and careful material selection to achieve the required
lattice conditions.

  Rare earths have been observed to form long metallic silicide
nanowires on Si (001)
surfaces\cite{PreinesbergeretAl98JPD,ChenetAl00APL,ChenetAl01MSEB,NogamietAl01PRB}.
A typical growth process\cite{ChenetAl00APL} begins with a flat Si
(001) substrate in UHV.  Submonolayer coverages of Er are then
evaporated onto the substrate.  Erbium forms a thermodynamically
stable silicide, ErSi$_{2}$, with a hexagonal crystal structure.  The
silicide grows so that the lowest lattice mismatch (-1.3\%) occurs
along the [11$\overline{2}$0] direction of the ErSi$_{2}$ and the
(110) direction of the Si.  This direction is the long axis of the
wires.  The transverse direction ([0001] for the ErSi$_{2}$;
(1$\overline{1}$0) for the Si) has a much larger (6.5\%) lattice
mismatch; as a result wire growth in that direction is strongly
disfavored.  The result after annealing is a large number of metallic
nanowires a few nm in width, one or two atomic layers in thickness,
and microns in length.  See Fig.~\ref{fig:silicide}.

  As with direct chemical synthesis, self-assembly techniques for
producing nanowires are relatively immature, but show signs of
great promise for certain classes of materials.  The requirements
for UHV conditions and extremely careful surface preparation are
stringent, however.  Research on more robust, fault-tolerant,
and engineerable self-assembly mechanisms is sure to be an active
field for some time.

\section{\small{CONCLUSIONS}}

  Metal nanowires are important both as components of future
technologies and as tools for examining fundamental science in metals
at the nm scale.  It is clear from the above that tremendous progress
has been made in the last twenty years on techniques for fabricating
such structures.  Nanowires with transverse dimensions below 20~nm
present a particular challenge.

  The techniques presented here have advantages and disadvantages that
make them well-suited to certain tasks.  Summarizing the main points:
\begin{itemize}
\item  Nanowires are only as well-defined as the metal material that
constitutes them.  For physically deposited metals small grain sizes
aid in the formation of narrow structures, and the surface physics
of the metal-substrate interface is critical in determining wire
morphology.  

\item Lithographic processes possess tremendous flexibility, but 
tend to be slow.  Lateral definition at extremely small scales is
very challenging, and true atomic resolution from scanned probe
methods requires UHV conditions and special substrate preparation.

\item Templating approaches have been very successful.  Subtractive
methods produce small numbers of extremely narrow wires at very
well-defined locations, given engineered (step-edge,
MBE-defined) templates.  

\item Additive templating using porous membranes to constrain
electrochemical metal growth produces very
large numbers of nanowires.  Addressing individual wires or
arraying them on a substrate is nontrivial, however.  Other
structures for templating (chemical
scaffolds, nanotubes) are also promising.

\item Direct chemical synthesis of metal nanowires is promising in
its infancy.  Encouraging results exist in a small number of material
systems, and much remains to be learned about growth mechanisms
and generalizability of techniques. 

\item Self-assembly is also a nascent approach.  Here, too,
the restrictions of particular material systems and processing
conditions present challenges that need to be more fully
investigated.
\end{itemize}

Nanowires promise to be a fruitful area of physics, materials science,
and chemistry research for the foreseeable future.  Advances in
nanoscale characterization techniques and computational approaches to
materials should ensure much continued progress in this exciting arena.

DN wishes to acknowledge and thank his nanowire collaborators
(R.L. Willett, L.N. Pfeiffer, K.W. West) and  support from
the DOE (DE-FG03-01ER45946) and the Robert A. Welch Foundation.

\bibliographystyle{rsp}

\bibliography{wirerev}

\begin{thebibliography}{100}

\bibitem{Ruitenbeek99Springer}
van Ruitenbeek, J.~M. 2000, in K.-H. Meiwes-Broer (Ed.), Metal Clusters on
  Surfaces: Structure, Quantum Properties, Physical Chemistry, Springer-Verlag,
  Berlin.

\bibitem{YansonetAl98Nature}
Yanson, A.~I., Bollinger, G.~R., van~den Brom, H.~E., Agrait, N., and van
  Ruitenbeek, J.~M. 1998, Nature, 395, 783.

\bibitem{CuietAl01APL}
Cui, Y., Lauhon, L.~J., Gudiksen, M.~S., Wang, J.~F., and Lieber, C.~M. 2001,
  Appl. Phys. Lett., 78, 2214.

\bibitem{PfeifferetAl93JCG}
Pfeiffer, L.~N., Stormer, H.~L., Baldwin, K.~W., West, K.~W., Goni, A.~R.,
  Pinczuk, A., Ashoori, R.~C., Dignam, M.~M., and Wegscheider, W. 1993, J.
  Cryst. Growth, 127, 849.

\bibitem{LiangetAl99APL}
Liang, C.~T., Simmons, M.~Y., Smith, C.~G., Ritchie, D.~A., and Pepper, M.
  1999, Appl. Phys. Lett., 75, 2975.

\bibitem{TansetAl97Nature}
Tans, S.~J., Devoret, M.~H., Dai, H.~J., Thess, A., Smalley, R.~E., Geerligs,
  L.~J., and Dekker, C. 1997, Nature, 386, 474.

\bibitem{Tour00ACR}
Tour, J.~M. 2000, Acc. Chem. Res., 33, 791.

\bibitem{Datta95}
Datta, S. 1995, Electronic transport in mesoscopic systems, Cambridge Studies
  in Semiconductor Physics and Microelectronic Engineering 3, Cambridge
  University, New York.

\bibitem{Imry97}
Imry, Y. 1997, Introduction to Mesoscopic Physics, Mesoscopic Physics and
  Nanotechnology, Oxford University, New York.

\bibitem{Feng91}
Feng, S. 1991, in B.~L. Altshuler, P.~A. Lee, and R.~A. Webb (Eds.), Mesoscopic
  phenomena in solids, Elsevier, Amsterdam, 107.

\bibitem{Giordano91}
Giordano, N. 1991, in B.~L. Altshuler, P.~A. Lee, and R.~A. Webb (Eds.),
  Mesoscopic phenomena in solids, Elsevier, Amsterdam, 131.

\bibitem{NatelsonetAl00SSC}
Natelson, D., Willett, R.~L., West, K.~W., and Pfeiffer, L.~N. 2000, Sol. State
  Comm., 115, 269.

\bibitem{ITRS01}
 2001, International Technology Roadmap for Semiconductors,
  http://public.itrs.net.

\bibitem{XiaetAl99CR}
Xia, Y., Rogers, J.~A., Paul, K.~E., and Whitesides, G.~M. 1999, Chem. Rev.,
  99, 1823.

\bibitem{Moreau88}
Moreau, W.~M. 1988, Plenum, New York.

\bibitem{Wong01}
Wong, A.~K. 2002, SPIE Press, Bellingham, Washington.

\bibitem{ReichmanisetAl01MS}
Reichmanis, E., Nalamasu, O., and Houlihan, F.~M. 2001, Macromol. Symp., 175,
  185.

\bibitem{BetzigetAl92Science}
Betzig, E. and Trautman, J.~K. 1992, Science, 257, 189.

\bibitem{KuwaharaetAl00ME}
Kuwahara, M., Nakano, T., Tominaga, J., Lee, M.~B., and Atoda, N. 2000,
  Microelect. Eng., 53, 535.

\bibitem{KrivaneketAl99U}
Krivanek, O.~L., Dellby, N., and Lupini, A.~R. 1999, Ultramicroscopy, 78, 1.

\bibitem{SmithetAl90PT}
Smith, H.~I. and Craighead, H.~G. 1990, Physics Today, 43, 24.

\bibitem{GrobmanetAl79IEEE}
Grobman, W.~D., Luhn, H.~E., Donohue, T.~P., Speth, A.~J., Wilson, A.,
  Hatzakis, M., and Chang, T. H.~P. 1979, IEEE J. Sol. State Circuits, SC-14,
  282.

\bibitem{RooksetAl87JVSTB}
Rooks, M.~J., Wind, S., McEuen, P., and Prober, D.~E. 1987, J. Vac. Sci. Tech.
  B, 5, 318.

\bibitem{Sharifi96}
Sharifi, F. 1996, Proc. of the Symp. on High Speed III-V Elect. for Wireless
  Appl., 316.

\bibitem{HarrisonetAl82APL}
Harrison, T.~R., Mankiewich, P.~M., and Dayem, A.~H. 1982, Appl. Phys. Lett.,
  41, 1102.

\bibitem{TiberioetAl93APL}
Tiberio, R.~C., Craighead, H.~G., Lercel, M., Lau, T., Sheen, C.~W., and
  Allara, D.~L. 1993, Appl. Phys. Lett., 62, 476.

\bibitem{ChenetAl93APL}
Chen, W. and Ahmed, H. 1993, Appl. Phys. Lett., 62, 1499.

\bibitem{YasinetAl01APL}
Yasin, S., Hasko, D.~G., and Ahmed, H. 2001, Appl. Phys. Lett., 78, 2760.

\bibitem{HatzoretAl01Science}
Hatzor, A. and Weiss, P.~S. 2001, Science, 291, 1019.

\bibitem{ProberetAl80APL}
Prober, D.~E., Feuer, M.~D., and Giordano, N. 1980, Appl. Phys. Lett., 37, 94.

\bibitem{Giordano80PRB}
Giordano, N. 1980, Phys. Rev. B, 22, 5635.

\bibitem{DynesetAl78PRL}
Dynes, R.~C., Garno, J.~P., and Rowell, J.~M. 1978, Phys. Rev. Lett., 40, 479.

\bibitem{HerzogetAl96PRL}
Herzog, A.~V., Xiong, P., Sharifi, F., and Dynes, R.~C. 1996, Phys. Rev. Lett.,
  76, 668.

\bibitem{ButkoetAl00PRL}
Butko, V.~Y., Ditusa, J.~F., and Adams, P.~W. 2000, Phys. Rev. Lett., 84, 1543.

\bibitem{GraybealetAl84PRB}
Graybeal, J.~M. and Beasley, M.~R. 1984, Phys. Rev. B, 29, 4167.

\bibitem{Quate97SS}
Quate, C.~F. 1997, Surf. Sci., 386, 259.

\bibitem{SohetAl01}
Soh, H.~T., Guarioni, K.~W., and Quate, C.~F. 2001, Kluwer Academic, New York.

\bibitem{Bai00}
Bai, C. 2000, Springer-Verlag, New York.

\bibitem{AbadaletAl98APA}
Abadal, G., Perez-Murano, F., Barniol, N., and Aymerich, X. 1998, Appl. Phys.
  A, 66, S791.

\bibitem{AvourisetAl98APA}
Avouris, P., Martel, R., Hertel, T., and Sandstrom, R. 1998, Appl. Phys. A, 66,
  S659.

\bibitem{PineretAl99Science}
Piner, R.~D., Zhu, J., Xu, F., Hong, S.~H., and Mirkin, C.~A. 1999, Science,
  283, 661.

\bibitem{RamspergeretAl01APL}
Ramsperger, U., Uchihashi, T., and Nejoh, H. 2001, Appl. Phys. Lett., 78, 85.

\bibitem{SohnetAl95APL}
Sohn, L.~L. and Willett, R.~L. 1995, Appl. Phys. Lett., 67, 1552.

\bibitem{BouchiatetAl96APL}
Bouchiat, V. and Esteve, D. 1996, Appl. Phys. Lett., 69, 3098.

\bibitem{HuetAl98JVSTB}
Hu, S., Hamidi, A., Altmeyer, S., Koster, T., Spangenberg, B., and Kurz, H.
  1998, J. Vac. Sci. Tech. B, 16, 2822.

\bibitem{SugawaraetAl95Science}
Sugawara, Y., Ohta, M., Ueyama, H., and Morita, S. 1995, Science, 270, 1646.

\bibitem{MajumdaretAl92APL}
Majumdar, A., Oden, P.~I., Carrejo, J.~P., Nagahara, L.~A., Graham, J.~J., and
  Alexander, L. 1992, Appl. Phys. Lett., 61, 2293.

\bibitem{KimetAl98JKPS}
Kim, J., Lee, H., Shin, Y., and Park, S. 1998, J. Korean Phys. Soc., 33, S84.

\bibitem{DavidssonetAl99ME}
Davidsson, P., Lindell, A., Makela, T., Paalanen, M., and Pekola, J. 1999,
  Micro. Eng., 45, 1.

\bibitem{DuboisetAl99SSE}
Dubois, E. and Bubbendorff, J.~L. 1999, Sol. State Elect., 43, 1085.

\bibitem{LydingetAl94APL}
Lyding, J.~W., Shen, T.~C., Hubacek, J.~S., J.~R.~Tucker, J., and Abeln, G.~C.
  1994, Appl. Phys. Lett., 64, 2010.

\bibitem{PalasantzasetAl99JAP}
Palasantzas, G., Ilge, B., Nijs, J.~D., and Geerligs, L.~J. 1999, J. Appl.
  Phys., 85, 1907.

\bibitem{MasuetAl94JVSTB}
Masu, K. and Tsubouchi, M. 1994, J. Vac. Sci. Tech. B, 12, 3270.

\bibitem{AdamsetAl96APL}
Adams, D.~P., Mayer, T.~M., and Swartzentruber, B.~S. 1996, Appl. Phys. Lett.,
  68, 2210.

\bibitem{BeckeretAl90PRL}
Becker, R.~S., Higashi, G.~S., Chabal, Y.~J., and Becker, A.~J. 1990, Phys.
  Rev. Lett., 65, 1917.

\bibitem{Bolland90PRL}
Bolland, J.~J. 1990, Phys. Rev. Lett., 65, 3325.

\bibitem{PalasantzasetAl99ME}
Palasantzas, G., Ilge, B., Rogge, S., and Geerlings, L.~J. 1999, Microelect.
  Eng., 46, 133.

\bibitem{SugimuraetAl93JJAP}
Sugimura, H., Uchida, T., Kitamura, N., and Masuhara, H. 1993, Jap. Journ.
  Appl. Phys. 2, 32, L553.

\bibitem{SnowetAl95Science}
Snow, E.~S. and Campbell, P.~M. 1995, Science, 270, 1639.

\bibitem{WangetAl95APL}
Wang, D.~W., Tsau, L.~M., Wang, K.~L., and Chow, P. 1995, Appl. Phys. Lett.,
  67, 1295.

\bibitem{SnowetAl96APL}
Snow, E.~S., Park, D., and Campbell, P.~M. 1996, Appl. Phys. Lett., 69, 269.

\bibitem{HeldetAl98PE}
Held, R., Heinzel, T., Studerus, P., and Ensslin, K. 1998, Physica E, 2, 748.

\bibitem{GiordanoetAl79PRL}
Giordano, N., Gilson, W., and Prober, D.~E. 1979, Phys. Rev. Lett., 43, 725.

\bibitem{FlandersetAl81JVST}
Flanders, D.~C. and White, A.~E. 1981, J. Vac. Sci. Tech., 19, 892.

\bibitem{WhiteetAl82PRL}
White, A.~E., Tinkham, M., Skocpol, W.~J., and Flanders, D.~C. 1982, Phys. Rev.
  Lett., 48, 1752.

\bibitem{GiordanoetAl89PRL}
Giordano, N. and Schuler, E.~R. 1989, Phys. Rev. Lett., 63, 2417.

\bibitem{HongetAl95JMMM}
Hong, K.~M. and Giordano, N. 1995, J. Mag. Mag. Mat., 151, 396.

\bibitem{NatelsonetAl00APL}
Natelson, D., Willett, R.~L., West, K.~W., and Pfeiffer, L.~N. 2000, Appl.
  Phys. Lett., 77, 1991.

\bibitem{Natelsonunpub}
Werder, D., Natelson, D., Willett, R.~L., Pfeiffer, L.~N., and West, K.~W.
  2001, unpublished.

\bibitem{NatelsonetAl01PRL}
Natelson, D., Willett, R.~L., West, K.~W., and Pfeiffer, L.~N. 2001, Phys. Rev.
  Lett., 86, 1821.

\bibitem{YunetAl00JVSTA}
Yun, W.~S., Kim, J., Park, K.~H., Ha, J.~S., Ko, Y.~J., Park, K., Kim, S.~K.,
  Doh, Y.~J., Lee, H.~J., Salvetat, J.~P., and Forro, L. 2000, J. Vac. Sci.
  Tech. A, 18, 1329.

\bibitem{SordanetAl01APL}
Sordan, R., Burghard, M., and Kem, K. 2001, Appl. Phys. Lett., 79, 2073.

\bibitem{ChoietAl00JVSTA}
Choi, S.~H., Wang, K.~L., Leung, M.~S., Stupian, G.~W., Presser, N., Morgan,
  B.~A., Robertson, R.~E., Abraham, M., King, E.~E., Tueling, M.~B., Chung,
  S.~W., Heath, J.~R., Cho, S.~L., and Ketterson, J.~B. 2000, J. Vac. Sci.
  Tech. A, 18, 1326.

\bibitem{HeathetAl97JPCB}
Heath, J.~R., Knobler, C.~M., and Leff, D.~V. 1997, J. Phys. Chem. B, 101, 189.

\bibitem{FritzscheetAl99APL}
Fritzsche, W., B{\"{o}}hm, K.~J., Unger, E., and K{\"{o}}hler, J.~M. 1999,
  Appl. Phys. Lett., 75, 2854.

\bibitem{BezryadinetAl00Nature}
Bezryadin, A., Lau, C.~N., and Tinkham, M. 2000, Nature, 404, 971.

\bibitem{BezryadinetAl97APL}
Bezryadin, A., Dekker, C., and Schmid, G. 1997, Appl. Phys. Lett., 71, 1273.

\bibitem{ZhangetAl00APL}
Zhang, Y. and Dai, H. 2000, Appl. Phys. Lett., 77, 3015.

\bibitem{KongetAl98Nature}
Kong, J., Soh, H.~T., Cassell, A.~M., Quate, C.~F., and Dai, H. 1998, Nature,
  395, 878.

\bibitem{ZhangetAl00CPL}
Zhang, Y., Frankln, N.~W., Chen, R.~J., and Dai, H. 2000, Chem. Phys. Lett.,
  31, 35.

\bibitem{Martin94Science}
Martin, C.~R. 1994, Science, 266, 1961.

\bibitem{Possin71Physica}
Possin, G.~E. 1971, Physica, 55, 339.

\bibitem{FleischeretAl75}
Fleischer, R.~L., Prince, P.~B., and Walker, R.~M. 1975, Univ. of California
  Press, Berkeley.

\bibitem{WilliamsetAl84RSI}
Williams, W.~D. and Giordano, N. 1984, Rev. Sci. Instr., 55, 410.

\bibitem{SchonenbergeretAl97JPCB}
Sch{\"{o}}nenberger, C., van~der Zaande, B. M.~I., Fokkink, L. G.~J., Henny,
  M., Schmid, C., Kr{\"{u}}ger, M., Bachtold, A., Huber, R., Birk, H., and
  Staufer, U. 1997, J. Phys. Chem. B, 101, 5497.

\bibitem{Despic89}
Despic, A. and Parkhutik, V.~P. 1989, in J.~O. Bockris, R.~E. White, and B.~E.
  Conway (Eds.), Modern Aspects of Electrochemistry, volume~26, Plenum, NY, Ch.
  6.

\bibitem{RoutkevitchetAl96IEEE}
Routkevitch, D., Tager, A.~A., Haruyama, J., Almawlawi, D., Moskovits, M., and
  Xu, J.~M. 1996, IEEE Trans. Elect. Dev., 43, 1646.

\bibitem{TonuccietAl92Science}
Tonucci, R.~J., Justus, B.~L., Campillo, A.~J., and Ford, C.~E. 1992, Science,
  258, 783.

\bibitem{NguyenetAl98JECS}
Nguyen, P.~P., Pearson, D.~H., tonucci, R.~J., and Babcock, K. 1998, J.
  Electrochem. Soc., 145, 247.

\bibitem{FasolkaetAl01ARMR}
Fasolka, M.~J. and Mayes, A.~M. 2001, Ann. Rev. Mat. Res., 31, 323.

\bibitem{ThurnAlbrechtetAl00Science}
Thurn-Albrecht, T., Schotter, J., Kastle, C.~A., Emley, N., Shibauchi, T.,
  Krusin-Elbaum, L., Guarini, K., Black, C.~T., Tuominen, M., and Russell,
  T.~P. 2000, Science, 290, 2126.

\bibitem{MartinetAl99AM}
Martin, B.~R., Dermody, D.~J., Reiss, B.~D., Fang, M.~M., Lyon, L.~A., Natan,
  M.~J., and Mallouk, T.~E. 1999, Adv. Mat., 11, 1021.

\bibitem{NicewarneretAl01Science}
Nicewarner-Pena, S.~R., Freeman, R.~G., Reiss, B.~D., He, L., Pena, D.~J.,
  Walton, I.~D., Cromer, R., Keating, C.~D., and Natan, M.~J. 2001, Science,
  294, 137.

\bibitem{PirauxetAl94APL}
Piraux, L., George, J.~M., Despres, J.~F., Leroy, C., Ferain, E., Legras, R.,
  Ounadjela, K., and Fert, A. 1994, Appl. Phys. Lett., 65, 2484.

\bibitem{LiuetAl95PRB}
Liu, K., Nagodawithana, K., Searson, P.~C., and Chien, C.~L. 1995, Phys. Rev.
  B, 51, 7381.

\bibitem{ZhangetAl98APL}
Zhang, Z.~B., Sun, X.~Z., Dresselhaus, M.~S., Ying, J.~Y., and Heremans, J.~P.
  1998, Appl. Phys. Lett., 73, 1589.

\bibitem{BachtoldetAl98ME}
Bachtold, A., Terrier, C., Kruger, M., Henny, M., Hoss, T., Strunk, C., Huber,
  R., Birk, H., Staufer, U., and Schonenberger, C. 1998, Micro. Eng., 42, 571.

\bibitem{YietAl99APL}
Yi, G. and Schwarzacher, W. 1999, Appl. Phys. Lett., 74, 1746.

\bibitem{GaoetAl01APA}
Gao, T., Meng, G.~W., Zhang, J., Wang, Y.~W., Liang, C.~H., Fan, J.~C., and
  Zhang, L.~D. 2001, Appl. Phys. A, 73, 251.

\bibitem{FasoletAl97APL}
Fasol, G. and Runge, K. 1997, Appl. Phys. Lett., 70, 2467.

\bibitem{HongetAl01SCience}
Hong, B.~H., Bae, S.~C., Lee, C.~W., Jeong, S., and Kim, K.~S. 2001, Science,
  294, 348.

\bibitem{Kimpriv01}
Kim, K.~S. 2002, private communication.

\bibitem{BraunetAl98Nature}
Braun, E., Eichen, Y., Sivan, U., and Ben-Yoseph, G. 1998, Nature, 391, 775.

\bibitem{PuntesetAl01Science}
Puntes, V.~F., Krishnan, K.~M., and Alivisatos, A.~P. 2001, Science, 291, 2115.

\bibitem{HuetAl99ACR}
Hu, J.~T., Odom, T.~W., and Lieber, C.~M. 1999, Acc. Chem. Res., 32, 435.

\bibitem{WagneretAl64APL}
Wagner, R.~S. and Ellis, W.~C. 1964, Appl. Phys. Lett., 4, 89.

\bibitem{Wagner70}
Wagner, R.~S. 1970, in A.~P. Levitt (Ed.), Whisker Technology, Wiley, New York.

\bibitem{JanaetAl01CC}
Jana, N.~R., Gearheart, L., and Murphy, C.~J. 2001, Chem. Comm., 7, 617.

\bibitem{GatesetAl00JACS}
Gates, B., Yin, Y., and Xia, Y. 2000, J. Am. Chem. Soc., 122, 12582.

\bibitem{MayersetAl01AM}
Mayers, B., Gates, B., Yin, Y., and Xia, Y. 2001, Adv. Mat., 13, 1380.

\bibitem{LiuetAl01AM}
Liu, S., Yue, J., and Gedanken, A. 2001, Adv. Mat., 13, 656.

\bibitem{WangetAl01CC}
Wang, Y., Zhang, L., Meng, G., Liang, C., Wang, G., and Sun, S. 2001, Chem.
  Commun., 2632.

\bibitem{JungetAl95APA}
Jung, T., Schlittler, R., Gimzewski, J.~K., and Himpsel, F.~J. 1995, Appl.
  Phys. A, 61, 467.

\bibitem{HimpseletAl97SRL}
Himpsel, F.~J., Jung, T., and Ortega, J.~E. 1997, Surf. Rev. Lett., 4, 371.

\bibitem{BatzilletAl98N}
Batzill, M., Sarstedt, M., and Snowdon, K.~J. 1998, Nanotechnology, 9, 20.

\bibitem{DekosteretAl99APL}
Dekoster, J., Degroote, B., Pattyn, H., Lagouche, G., Vantomme, A., and
  Degroote, S. 1999, Appl. Phys. Lett., 75, 938.

\bibitem{ZachetAl00Science}
Zach, M.~P., Ng, K.~H., and Penner, R.~M. 2000, Science, 290, 2120.

\bibitem{ChenetAl01MSEB}
Chen, Y., Ohlberg, D. A.~A., and Williams, R.~S. 2001, Mat. Sci. Eng. B, 87,
  222.

\bibitem{PetroffetAl01PT}
Petroff, P.~M., Lorke, A., and Imamoglu, A. 2001, Phys. Today, 54, 46.

\bibitem{ChenetAl00APL}
Chen, Y., Ohlberg, D. A.~A., Medeiros-Ribeiro, G., Chang, Y.~A., and Williams,
  R.~S. 2000, Appl. Phys. Lett., 76, 4004.

\bibitem{PetroffetAl94SLM}
Petroff, P.~M. and DenBaars, S.~P. 1994, Superlatt. Microstr., 15, 15.

\bibitem{PreinesbergeretAl98JPD}
Preinesberger, C., Vandre, S., Kalka, T., and Dahne-Prietsch, M. 1998, J. Phys.
  D, 31, L43.

\bibitem{NogamietAl01PRB}
Nogami, J., Liu, B.~Z., Katkov, M.~V., Ohbuchi, C., and Birge, N.~O. 2001,
  Phys. Rev. B, 63, 233305.

\end{thebibliography}


\end{document}